# Electronic correlation determining correlated plasmons in Sb-doped Bi$_2$Se$_3$

P. K. Das[1,*], T. J. Whitcher[1,2,†], M. Yang[4,§], X. Chi[1,3], Y. P. Feng[3], W. Lin[5], J. S. Chen[5], I. Vobornik[6], J. Fujii[6], K. A. Kokh[7,8], O.E. Tereshchenko[8,9], C. Z. Diao[1], Jisoo Moon[10], Seongshik Oh[10], A. H. Castro-Neto[2,3], M. B. H Breese[1,3], A. T. S. Wee[2,3], and A. Rusydi[1,2,3,11,12,‡]

[1] *Singapore Synchrotron Light Source, National University of Singapore, 5 Research Link, 117603, Singapore*

[2] *Centre for Advanced 2D Materials, National University of Singapore, 6 Science Drive 2, 117546, Singapore*

[3] *Department of Physics, National University of Singapore, 2 Science Drive 3, 117576, Singapore*

[4] *Institute of Materials Research and Engineering, A∗STAR (Agency for Science, Technology and Research), 2 Fusionopolis Way, 138634, Singapore*

[5] *Department of Materials Science and Engineering, National University of Singapore, 117575, Singapore*

[6] *Istituto Officina dei Materiali (IOM) - CNR, Laboratorio TASC, Area Science Park, S.S.14, Km 163.5, I-34149 Trieste, Italy*

[7] *V.S. Sobolev Institute of Geology and Mineralogy, Novosibirsk, 630090 Russia*

[8] *Novosibirsk State University, Novosibirsk, 630090 Russia*

[9] *A.V. Rzhanov Institute of Semiconductor Physics, Novosibirsk, 630090 Russia*

[10] *Department of Physics & Astronomy, Rutgers, The State University of New Jersey, 136 Frelinghuysen Road, Piscataway, New Jersey 08854, USA*

[11] *NUSSNI-NanoCore, National University of Singapore, 117576, Singapore*

[12] *NUS Graduate School for Integrative Sciences and Engineering, 117456, Singapore*

Corresponding authors: [*]das@nus.edu.sg, [†]c2dwtj@nus.edu.sg, [‡]phyandri@nus.edu.sg

**Abstract:** Electronic correlation is believed to play an important role in exotic phenomena such as insulator-metal transition, colossal magneto resistance and high temperature superconductivity in correlated electron systems. Recently, it has been shown that electronic correlation may also be responsible for the formation of unconventional plasmons. Herewith, using a combination of angle-dependent spectroscopic ellipsometry, angle resolved photoemission spectroscopy and Hall measurements all as a function of temperature supported by first-principles calculations, the existence of low-loss high-energy correlated plasmons accompanied by spectral weight transfer, a fingerprint of electronic correlation, in topological insulator $(Bi_{0.8}Sb_{0.2})_2Se_3$ is revealed. Upon cooling, the density of free charge carriers in the surface states decreases whereas those in the bulk states increase, and that the newly-discovered correlated plasmons are key to explaining this phenomenon. Our result shows the importance of electronic correlation in determining new correlated plasmons and opens a new path in engineering plasmonic-based topologically-insulating devices.

**Introduction:** The electronic structure of a three-dimensional topological insulator (TI) consists of novel topological surface states originating from the massless Dirac fermions and insulating bulk states [1-5]. The time reversal symmetry protected surface states are gapless, while the Fermi level lies in-between the gapped bulk band structure, resulting in an insulating bulk and a conducting surface. Topological insulators have attracted a flurry of research activities not only for their fundamental importance, but they are also being proposed for various real life applications, including field effect transistors [6-8], next generation quantum computers [9, 10], high-speed opto-spintronic applications [11], and $Bi_2Se_3$-assited membrane crystallization, which can be exploited in various applications such as water desalination and minerals recovery [12]. In recent years, the field of plasmonics has attracted considerable attention because of its novel technological applications including confining light into sub-wavelength dimensions [13-16]. As such, efforts have been made to study the Dirac [17-21] and surface plasmons [20, 22-25] at low energies that have been shown to occur in TIs with the aim of producing highly efficient, high-speed plasmonic devices [26-28]. In fact, $Bi_2Se_3$ has already been demonstrated as teraheartz photodetector [29-31] exploiting the plasmonic modes arising from its topological surface states and thermoplasmonic devices [32].

Recently, a new type of plasmon, correlated plasmons, has been theoretically proposed to occur due to electronic correlation in strongly-correlated materials [33]. Indeed, these correlated plasmons have been observed in the series of Mott-like insulating oxides [34, 35]. Correlated plasmons differ from conventional plasmons in that they arise from the collective oscillation of correlated electrons as opposed to the collective oscillation of free charges. Due to the electron-electron interaction, correlated plasmons may have a positive, low real part of the dielectric function and can readily couple with free-space photons [33, 34, 36]. Because of this, they also have significantly lower loss than conventional plasmons, which could have benefits in devices utilizing topologically insulating materials. This motivates us to search and explore electronic correlation and new plasmons in TIs.

In this work, we report unusual effects of temperature change on the electronic and optical properties of the topological insulator $(Bi_{0.8}Sb_{0.2})_2Se_3$. Using a combination of angle-dependent spectroscopic ellipsometry, angle resolved photoemission spectroscopy and Hall measurements all as a function of temperature supported by first-principles calculations, we discover novel low-temperature high-energy correlated plasmons accompanied by anomalous spectral weight transfer in topological insulators. Upon cooling, the spectral weight transfer of free charge carriers occurs between the surface and the bulk states, which is due to electronic correlation and the scattering of surface state free charge carriers from the correlated plasmons at lower temperatures.

**Experimental methods and calculations:**

**Sample preparation and ARPES.** High quality single crystalline sample of $(Bi_{0.8}Sb_{0.2})_2Se_3$ is grown by Bridgman method [37]. The single crystal is cleaved in UHV by standard Kapton-tape method to obtain good quality surface for the ARPES measurement. The ARPES experiment is performed at CNR-IOM APE beamline [38] at Elettra Sincrotrone Trieste. The Fermi surface maps are performed using Scienta DA30 electron analyzer without rotating the sample. The data are collected at liquid nitrogen temperature (T = 78 K), and at chamber base pressure better than $1 \times 10^{-10}$ mbar. The temperature dependent ARPES and XPS measurements are also performed at soft X-ray and ultraviolet (SUV) beamline of Singapore Synchrotron Light Source [39]. The ARPES measurements are carried out using laboratory Helium lamp source (photon energy = 21.21 eV), while XPS measurements are carried out using a laboratory Al $K\alpha$ (photon energy = 1486.6 eV) X-ray source. **Spectroscopic Ellipsometry.** Spectroscopic ellipsometry measurements are carried out using a variable-angle spectroscopic ellipsometer (V- VASE, J.A. Woollam Co.) with a rotating analyzer and compensator at the Singapore Synchrotron Light Source (SSLS). The spectroscopic ellipsometry is a non-destructive photon-in photon-out technique, therefore it does not have charging issue and can be used to probe simultaneously loss function and complex dielectric function in correlated electron systems. Also, spectroscopic ellipsometry is not as surface sensitive as photoelectron spectroscopy techniques, where the light penetration depth is tens or even hundreds of nm as shown by the attenuation length in Supplementary Fig. S2 [40], which is calculated from the complex dielectric function in Fig. 2. The measurements are taken in the energy range of 0.6 – 6.0 eV whilst the sample was inside an UHV cryostat with a base pressure of $10^{-8}$ mbar. The measurements are taken at angles of 68°, 70° and 72°, which are limited in range by the UHV windows. Measurements are taken at a range of temperatures from 77 K to 475 K at an angle of 70°. As the sample is a thick single (>10 μm) crystal, the complex dielectric function can be determined through the best fits to the output data $\psi$ and $\Delta$, which are shown in Supplementary Fig. S1(a) and S1(b) respectively [40]. As single crystal $(Bi_{0.8}Sb_{0.2})_2Se_3$ is anisotropic, the spectroscopic ellipsometer is used in the Mueller-Matrix mode with a rotational sample stage. Further details can be found in Refs. [41] and [42]. Using the W-VASE analysis program, a model of the sample is created, which includes surface effects such as roughness and contamination. As the process of spectroscopic ellipsometry is a self-normalising technique, the complex dielectric function of the sample can be determined through the best fits to the output data $\psi$ and $\Delta$ [41-44]. **Hall Measurements.** The Hall effect measurement is performed in a physical properties measurement system (PPMS), where a bar-shape device was prepared, and silver paste is used as the electrodes. **First-Principle Calculations**. The first-principles calculations are performed using density-functional theory based Vienna ab initio simulation package (VASP) [45, 46] with the Perdew-Burke Ernzerhof (PBE) functional and projector augmented wave (PAW) potentials [47]. In all calculations, the cutoff energy for the electronic planewave expansion is set to 500 eV, and spin-orbital coupling effect is included. The criterion for electronic energy convergence is set to $1.0\times10^{-6}$ eV. The lattice constants of the pristine $Bi_2Se_3$ bulk are fixed to experimental values (a=b=4.14 Å and c=28.64 Å) [48, 49], while the atomic positions are optimized with the van der Waals correction (DFT-D3) until the force is smaller than 0.01 eV/ Å [50]. The topological surface states are calculated using six quintuple layers (QL) of the $Bi_2Se_3$ and 20 Å vacuum layers normal to the surface. The two Bi atoms in the top and bottom QL layers are substituted by the Sb atoms to simulate the Sb doping effect, as shown in Figure 5. $\Gamma$-centered $6 \times 6 \times 6$ and $9 \times 9 \times 1$ $k$-point meshes are used to sample the first Brillouin zone (BZ) of the $Bi_2Se_3$ primitive cell and the surface slabs, respectively.

**Results:** The high-resolution ARPES results of $(Bi_{0.8}Sb_{0.2})_2Se_3$ presented in Fig. 1(a) show the 3D overview of the Fermi surface and band dispersion along the $\Gamma - K$ direction. A well-defined Dirac cone, the hallmark of topological insulators, is clearly observed along with a portion of the bulk conduction band, which lies below the Fermi level. These extra carriers may be the result of selenium deficiency and the presence of other impurities in the system [51-53]. However, by means of Sb substitution, we are able to control the Fermi level in this system. With a 20% Sb substitution, the Dirac cone is located 0.27 eV below the Fermi level, which is smaller when compared to undoped $Bi_2Se_3$ (in undoped $Bi_2Se_3$ the Dirac point is $0.30 - 0.40$ eV below the Fermi level) [51, 54, 55]. It is important to note that even though Bi and Sb are isovalent, variable Sb doping controls the defect induced bulk carrier density in this system. (For the discussions on possible surface contamination, we refer to supplementary Fig. S3 and supplementary note [40]). Fig. 1(b) presents the constant energy cuts at binding energy intervals of 0.1 eV, starting from the Fermi surface down to a binding energy of 0.4 eV. This gives a clearer picture of the overall electronic structure of the Sb-doped $Bi_2Se_3$ system. These measurements are performed using a photon energy of 30 eV. Photon energy dependent measurements are shown in Fig. 1(c-e) presenting the bandstructure along the $\Gamma - K$ direction at photon energies of $h\nu = 30$ eV, 40 eV and 55 eV, respectively. Photon energy dependent measurement disentangles the bulk and surface features; the Dirac surface state is prominent through all photon energy values while the inner conduction band disperses strongly with probe energy, which is reminiscent of its bulk character.

From Fig. 1, the bottom of the conduction band has a binding energy of ~0.16 eV and the Dirac point is located at 0.27 eV below the Fermi level ($E_F$), with a Fermi wave vector $k_F \sim \pm$ 0.08 Å$^{-1}$. The bulk free carrier density of our system can now be estimated using these ARPES results. We may use the approximation of parabolic bands to find the free carrier density in the bulk of the system [56, 57]: $E_{CB} = (\hbar^2/2m^*)(3\pi^2 n_{BD})^{2/3}$ (2), where $E_{CB}$ is the binding energy of the bottom of conduction band, $n_{BD}$ is the bulk free carrier density and m* is the effective free charge carrier mass. Considering an effective mass ($m^*$) value of ≈ 0.15 $m_e$, typical for a Bi$_2$Se$_3$ system [58], a bulk free carrier density of 1.69×10$^{19}$ cm$^{-3}$ is obtained at 77 K.

Figures 2(a) and 2(b) show the real and imaginary parts of the dielectric function respectively, modelled from spectroscopic ellipsometry data taken over a range of temperatures from 475 K down to 77 K. Interestingly, we find anomalous spectral weight transfer in a broad energy range, from high energy to low energy upon cooling (as further discussed later). Spectral weight transfer is a fingerprint of electronic correlation [59-62]. There is a clear change in both parts of the complex dielectric function as the (Bi$_{0.8}$Sb$_{0.2}$)$_2$Se$_3$ is cooled, especially from 300 K to 250 K. There is also an edge that occurs in the $\varepsilon_2$ spectra at 0.95 eV for all temperatures that gets sharper as the sample is cooled.

Spectroscopic ellipsometry is a powerful tool used to search for plasmonic activity in correlated electron systems [34]. For photons below X-ray energy levels, the crystal momentum is much higher than the momentum transfer ($q$), therefore $q$ is finite but approaches zero. In this limit, the distinction between the longitudinal and transverse $\varepsilon(\omega)$ vanishes, which allows spectroscopic ellipsometry to probe both optical and plasmonic properties of materials in the low-$q$ limit through the loss function, which is calculated from the complex dielectric function [34, 63, 64].

Our main finding is a correlated plasmon at ~1 eV. Figure 2c shows a comparison of the loss function and $\varepsilon_2$ for all temperatures measured between 475 K and 77 K in the spectral region 0.6 eV – 1.4 eV. Interestingly, the loss function peak at ~1 eV occurs upon cooling when the spectral weight of $\varepsilon_2$ at higher energy is reduced [as shown in Fig. 2(b) and discussed later in Fig. 3(c)]. Such a spectral weight transfer is a fingerprint of electronic correlation, which is mainly responsible for the correlated plasmons [33, 34]. We note that this loss function peak is blue-shifted from the optical excitation in $\varepsilon_2$ at 0.95 eV, while $\varepsilon_1$ is still positive due to electron-electron interaction. The significant change in the spectral weight of $\varepsilon_2$ as the $(Bi_{0.8}Sb_{0.2})_2Se_3$ is cooled is further highlighted by the difference in the complex dielectric function with temperature shown in Supplementary Fig. S1(c) and S1(d) [40]. This is the first time that correlated plasmons have been detected in topological insulators.

In order to find the origin of correlated plasmons, we quantitively explore the spectral weight transfer that can be seen in the topological insulator's electronic response to external electromagnetic fields with temperature. In particular, we look at the optical conductivity, which is related to $\varepsilon_2$ by: $\sigma_1 = \omega\varepsilon_2/4\pi$ (3), where $\sigma_1$ is the real part of the complex conductivity and $\omega$ is the photon frequency.

Figure 3(a) shows the optical conductivity of the single crystal $(Bi_{0.8}Sb_{0.2})_2Se_3$ for each of the temperatures measured. The charge conserving $f$-sum rule: $\int_0^\infty \sigma_1(\omega)d\omega = \pi n_e e^2/2m_e$ (4), links the integral of the optical conductivity across the whole spectrum to the free-charge carrier density, $n_e$ [34, 63]. Therefore, the integration of a part of the spectral region between $E_1$ and $E_2$, given by: $W = \int_{E_1}^{E_2} \sigma_1(E)dE$ (5), is proportional to the number of free charge carriers within that spectral region. By analyzing the change of *W* over the spectral range 0.6 eV – 6.3 eV for each of the temperatures measured, as shown in Fig. 3(b), we can gain an insight into the behavior of the free charge carriers as the $Bi_2Se_3$ sample is cooled [34, 63].

There is a slight increase in the spectral weight as the sample is cooled from 475 K to 300 K, before a sharp decrease down to 250 K. There is then a gradual increase back up to $2.7 \times 10^4$ $(\Omega cm)^{-1}$ at 77 K. This indicates a drastic loss of electrons with energies between 0.6 eV – 6.3 eV between 300 K and 250 K. The energies required to shift the electrons outside of this spectral range are of the order of eV's and this rules out thermal energy transfer as the energies associated with temperatures below 500 K are too small (<43 meV). Therefore, the extra energy gained or lost must come from potential energy transfer i.e. electron-electron correlations. The drop in the conductivity, and thus electron density, that occurs at temperatures of 250 K and below also coincides with the appearance of the correlated plasmons seen in Fig. 2(c) at this temperature. It is a subject of a future study as to why this occurs below 250 K, however it may be worth to note that a previous theoretical study on a topological insulator $BiTlSe_2$ has shown that electron-phonon interactions became significant below 250 K and the system might enter the topologically insulating phase [65].

The optical conductivities for each temperature shown in Fig. 3(a) are divided into three spectral regions, which are then integrated across each region to give the results shown in Fig. 3(c). The positive region of the y-axis indicates an increase in the overall energy of the spectral range, whilst the negative region indicates a decrease in the overall energy. The lower energy region shows only a minor increase in spectral weight as the sample is cooled whereas the mid-energy region, between 1.60 eV and 2.75 eV, shows a larger increase. The high-energy region initially shows an increase in spectral weight followed by a massive decrease between 300 K and 250 K.

A smaller decrease is also present in the mid-energy region between these temperatures, but a slight increase is seen in the low-energy region. As the largest loss of electrons occurs within the high-energy region, it appears that electrons are gaining energies of the order of eV's to move outside of the upper measured limit of the spectral range rather than losing energy. As previously stated, this must be due to an increase in potential energy from long-range electron-electron correlations and the formation of correlated plasmons, which can only happen if electronic screening is being enhanced at these lower temperatures. It is found that although the integrated conductivity (i.e. the electron density) increases somewhat as the temperature is lowered from 100 K to 77 K, there is still an overall loss in that spectral region. There is also a subsequent increase of integrated conductivity and electron density within the middle region due to the main spectral features, which are not directly related to the correlated plasmons, but which contribute to the overall increase seen in Fig. 3b.

Using the optical conductivities from Fig. 3, and the charge carrier mobility, $\mu_e$, of $(Bi_{0.8}Sb_{0.2})_2Se_3$ in the following equation: $\sigma = n_e e \mu_e$ (6), the free charge carrier density, $n_e$, of both the bulk and surface states can be calculated. Note that the electron mobility of $(Bi_{0.8}Sb_{0.2})_2Se_3$ may be different from pure $Bi_2Se_3$ due to Sb doping [66].

Figure 4(a) shows the Hall measurements of the $(Bi_{0.8}Sb_{0.2})_2Se_3$ at 78 K, where the electron mobility was determined to be 1400 cm$^2$/Vs and the bulk carrier density is $1.67 \times 10^{19}$ cm$^{-3}$, which is very close to the calculated bulk free carrier density from the ARPES measurements of $1.69 \times 10^{19}$ cm$^{-3}$. This can then be compared with the electron mobility and carrier density from pure $Bi_2Se_3$ [66]. The carrier density of $Bi_2Se_3$ at 77 K is slightly lower at $1.3 \times 10^{19}$ cm$^{-3}$, however, the mobility is 1380 cm$^2$/Vs, which is very close to the $(Bi_{0.8}Sb_{0.2})_2Se_3$ measurement. Using this mobility, the measured optical conductivity and equation *(6)*, the total free charge carrier density is calculated to be $1.20 \times 10^{20}$ cm$^{-3}$ at 78 K.

By comparing this with the calculations of the electron density in the conduction band from the ARPES measurements, it can be seen that there is almost an order of magnitude difference between these estimates. This is because the electron density extracted from the optical conductivity is the total free electron density for both the bulk and the surface, whereas the electron density from the ARPES measurements is from the conduction bands in the supposedly insulating bulk of the sample. However, the bulk states can be considered a bad conductor in most cases or more accurately as a weaker conductor than the surface states, because in reality, few topologically insulating samples rarely achieve a truly insulating bulk due to the Mott criterion and Ioffe-Regel criterions [56]. With these calculated electron densities, the percentage of carriers from the surface states that contribute to the overall conduction is estimated at 85.9%.

At 300 K, the electron mobility of $(Bi_{0.8}Sb_{0.2})_2Se_3$ is determined to be 993 cm$^2$/Vs from the Hall measurements as shown in Fig. 4(b) and the bulk carrier density of $1.56 \times 10^{19}$ cm$^{-3}$ for our sample. The carrier density is $Bi_2Se_3$ at 300 K is again slightly lower at $1.4 \times 10^{19}$ cm$^{-3}$ and the mobility is measured to be 880 cm$^2$/Vs, which is also smaller than the Sb-doped $Bi_2Se_3$ [66]. By using equation *(6)* again with the measured mobility and conductivity, the total free charge carrier density is calculated to be $1.70 \times 10^{20}$ cm$^{-3}$ at 300 K. This is a significant increase from the charge carrier density at 78 K. By using the bulk carrier density from the Hall measurements, the percentage of free carriers coming from the surface states of room temperature Sb-doped $Bi_2Se_3$ is estimated to be 90.8%, which is higher than the estimate at 78 K.

This result shows that by decreasing the temperature of the TI, the overall free charge carrier density is lowered but the bulk carrier density has increased. This can only happen if free charge carriers are being scattered from the surface states to the bulk states. However, it is well known that the scattering of surface state charge carriers in topological insulators is extremely limited. This is because back-scattering from non-magnetic impurities is prohibited by time-reversal symmetry [4]. Other methods of scattering are also negligible, as phonons are too weak for electron scattering [67] and surface Dirac plasmons have energies of the order of 10 meV (in the THz regime) which are not enough to cause significant scattering [18]. However, since the correlated plasmons seen in Fig. 2 have energies of the order of 1 eV, this is enough to induce electron scattering from the surface states to the bulk as seen in other 2D materials such as graphene [68]. The coupling of electrons with these low-temperature correlated plasmons is the most likely reason behind the transfer of free charge carriers from the surface to the bulk conduction bands as the temperature is lowered.

Our analysis is further supported by first-principles calculations based on the density functional theory. The calculated band structure of Sb-doped $Bi_2Se_3$ surface slab is shown in Fig. 5(a), along with the projected contribution of surface QLs (red color) and central QLs (blue color). We can see that with the Sb doping, the surface Dirac cone is shifted slightly below the Fermi level, indicating that the $Bi_2Se_3$ stays *n*-doped after the Sb incorporation. These excess electrons might induce stronger Columbic interaction among the electrons, resulting in altered plasmonic properties. The relativistic spin-orbit coupling effect is self consistently taken into account in the calculations. It is noteworthy that the spin-orbit coupling plays an important role in the formation of electronic bands of TIs, and significantly affect the optical properties as seen in related materials [69, 70]. From the band structure, the contribution from the central QLs (bulk bands) is noticeable in the valence bands near the Fermi level. It hybridizes with the surface QLs (surface bands), forming the lower Dirac cone. This infers that with Sb doping, the electronic states from the bulk region extend to the surface layers, consistent with the experimental observation.

The hybridization between surface and bulk carriers is evidenced by the visualized partial charge density. As Fig. 5(b) shows, the carriers in the Dirac cone below the Fermi level are mainly from the surface QLs, but the contribution from the bulk region can be seen as well. The projected density of states (PDOSs) further corroborates the above observations. The contribution to the DOSs near the Fermi level is found predominantly from $p_z$ orbitals of Bi atoms in the surface QLs, which are hybridized with those from bulk QLs as shown in Fig. 5(c) and (d). This is in contrast with that of pristine $Bi_2Se_3$ (seen in Supplementary Fig. S5 [40]), where only the electrons in the surface QLs contribute to the surface states. Thus, more conducting electrons are found near the Fermi level in the $Bi_{2-x}Sb_xSe_3$ than those in the pristine $Bi_2Se_3$. It may be worth to note that it has been theoretically studied by Zhu et al. [36] that local field effects such as those due to spin-orbit coupling may play an important role that causes the spectral weight transfer thus generating unconventional plasmons at low temperatures.

**Conclusions:** In summary, by lowering the temperature of $(Bi_{0.8}Sb_{0.2})_2Se_3$ below 250 K, we have discovered the existence of low-loss correlated plasmons at high energy due to the electronic correlation. This is achieved through the determination of simultaneously complex dielectric function, loss function, and electronic structure and dispersion of the material as a function of temperature using spectroscopic ellipsometry, ARPES, Hall measurements and first-principle calculations. We reveal spectral weight transfer of free charge carrier density from the surface to the bulk as the temperature decreases. This spectral weight transfer in the topological conductivity of the material is due to electrons in the surface states scattering into the bulk states from the high-energy correlated plasmons at low temperatures. By controlling the correlated plasmonic behavior in Sb-doped Bi$_2$Se$_3$ through temperature changes, the topological conductivity of TIs can be manipulated, which may lead to a new advanced control over future plasmonic devices.

**Acknowledgment:** We thank Z. Li, E. Chew, H. Miao, W. Wong, W. Zaw, C Lim and T. C. Asmara for technical support. This work is supported by 2015 PHC Merlion Project, MOE-AcRF Tier-2 (Grants No. MOE2017-T2-1-135, MOE2015-T2-1-099, No. MOE2015-T2-2-065, and No. MOE2015-T2-2-147), the Singapore National Re- search Foundation under its Competitive Research Funding (Grants No. NRF-CRP 8-2011-06 and No. R-398-000-087- 281), and MOE-AcRF Tier 2 (Grants No. R-144-000-398-114, No. R-144-000-368-112, No. R- 144-000-346-112, and No. R-144-000-364-112, and R-144-000-423-114). The authors would also like to acknowledge the Singapore Synchrotron Light Source (SSLS) for providing the facility necessary for conducting the research. The Laboratory is a National Research Infrastructure under the National Research Foundation Singapore via NUS Core Support Grant C-380-003-003-001. Centre for Advanced 2D Materials and Graphene Research Centre at the National University of Singapore are acknowledged for providing the computing resource. The work at CNR-IOM APE beamline has been performed in the framework of the nanoscience foundry and fine analysis (NFFA-MIUR Italy Progetti Internazionali) facility. K.A.K and O.E.T. acknowledge financial support by the Russian Science Foundation (project n. 17-12-01047), in part of crystal growth and structural characterization. P. K. D., T. J. W., and M. Y. contributed equally to this work.

**Figure and figure captions:**

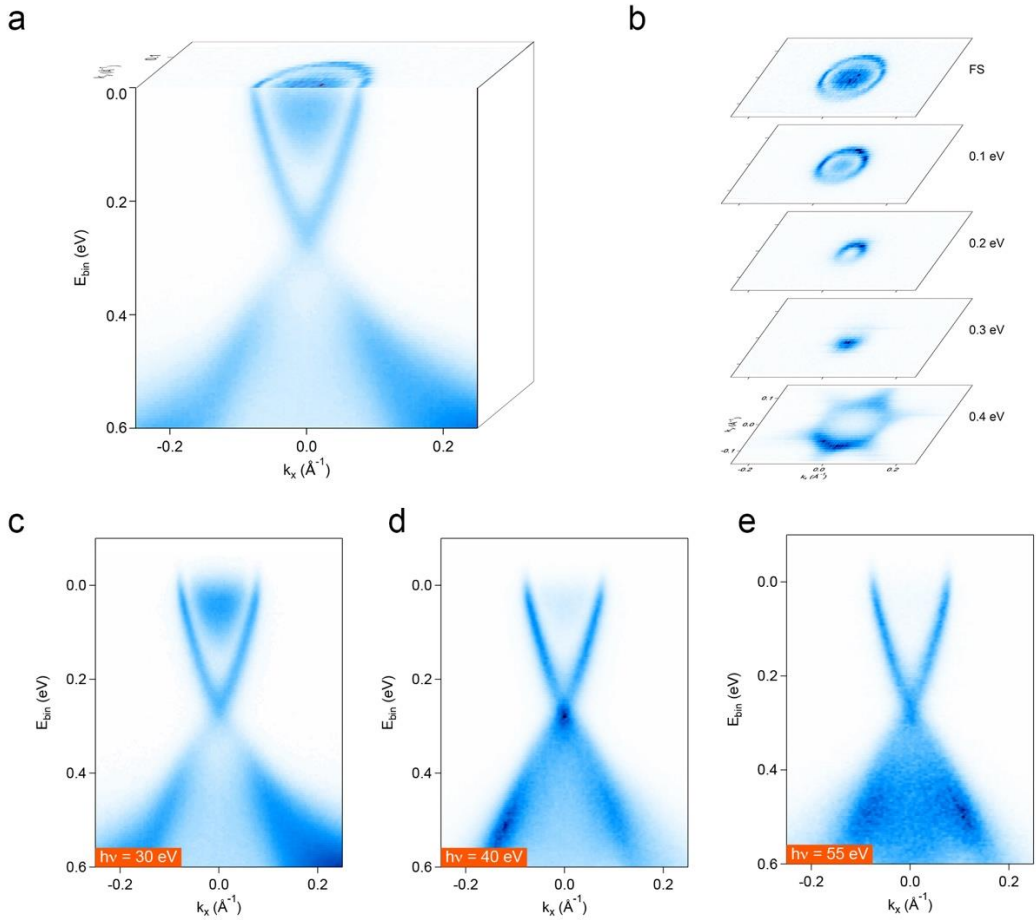

**Fig. 1.** Electronic structure of $(Bi_{0.8}Sb_{0.2})_2Se_3$. (a) The electronic structure of Sb-doped $Bi_2Se_3$ measured by high resolution ARPES ($h\nu$ = 30 eV, T = 78 K); the front surface of the 3D box presents the band structure along the $\Gamma - K$ direction while the top surface presents the Fermi surface ($k_x$ vs. $k_y$). (b) The constant energy cuts starting from Fermi surface (top) and at 0.1 eV, 0.2 eV, 0.3 eV and 0.4 eV (bottom) binding energies. (c-e) The band dispersion probed by three different photon energies of 30, 40, and 55 eV, respectively.

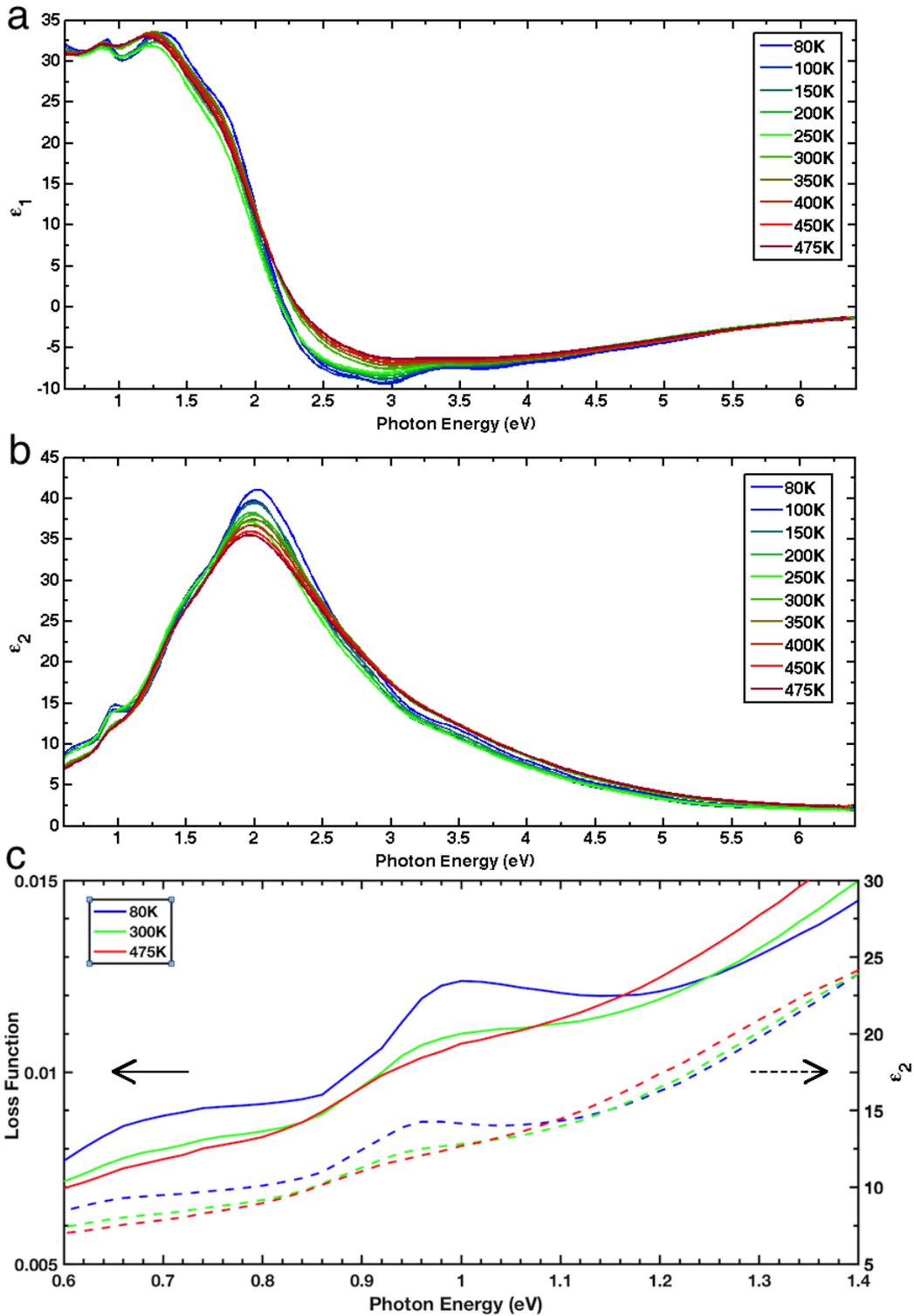

**Fig. 2.** Spectroscopic Ellipsometry of $(Bi_{0.8}Sb_{0.2})_2Se_3$. (a) The real and (b) imaginary parts of the dielectric function of $(Bi_{0.8}Sb_{0.2})_2Se_3$ modelled from spectroscopic ellipsometry data as a function

of temperature. (c) Comparison of the loss function $Im(-1/\epsilon)$ (solid line), and the imaginary part of the dielectric function, $\epsilon_2$ (dashed line) for all temperatures.

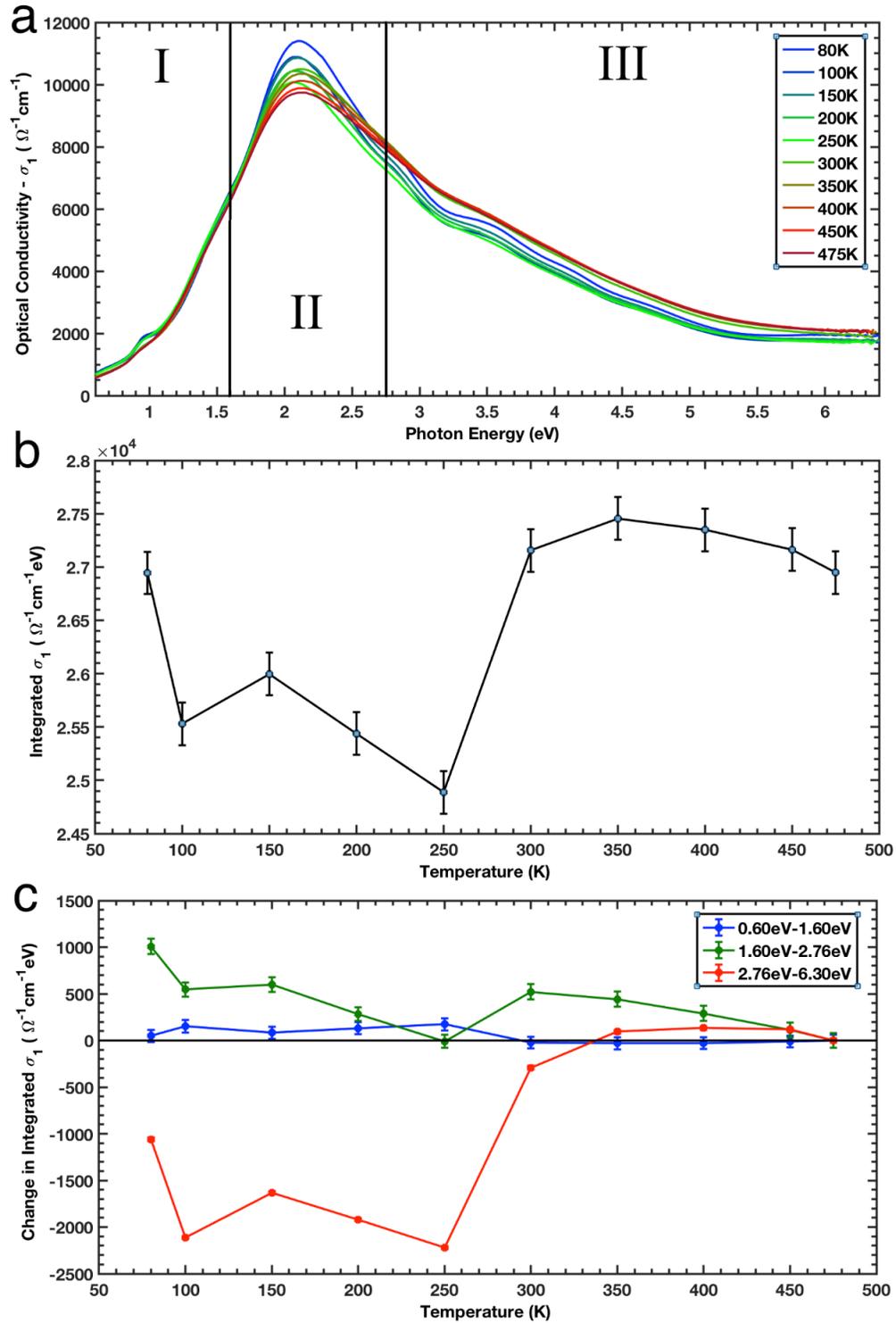

**Fig. 3.** Spectral weight change with temperature. (a) The optical conductivity of $(Bi_{0.8}Sb_{0.2})_2Se_3$ as a function of temperature and split into three different spectral regions. (b) The optical conductivity of $(Bi_{0.8}Sb_{0.2})_2Se_3$ integrated over the whole spectrum as a function of temperature. (c) The change in integrated conductivity of three different spectral regions as a function of temperature.

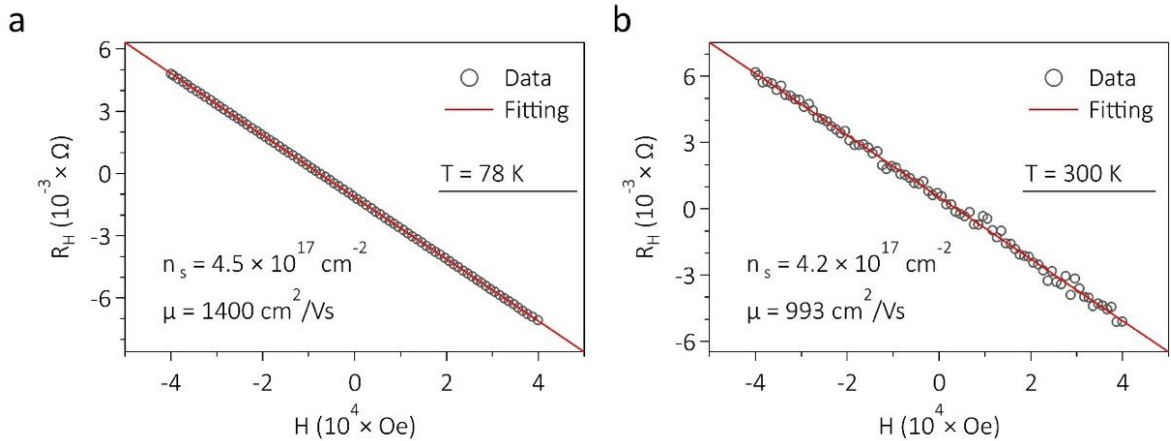

**Fig. 4.** Electron mobility of $(Bi_{0.8}Sb_{0.2})_2Se_3$. Hall measurements of the $(Bi_{0.8}Sb_{0.2})_2Se_3$ sample to determine the carrier density and the electron mobility at (a) 78 K where the carrier density and the electron mobility are $n_s = 4.5 \times 10^{17}$ cm$^{-2}$ and $\mu_e = 1400$ cm$^2$/Vs respectively, and (b) at 300 K, which are $n_s = 4.2 \times 10^{17}$ cm$^{-2}$ and $\mu_e = 993$ cm$^2$/Vs respectively. .

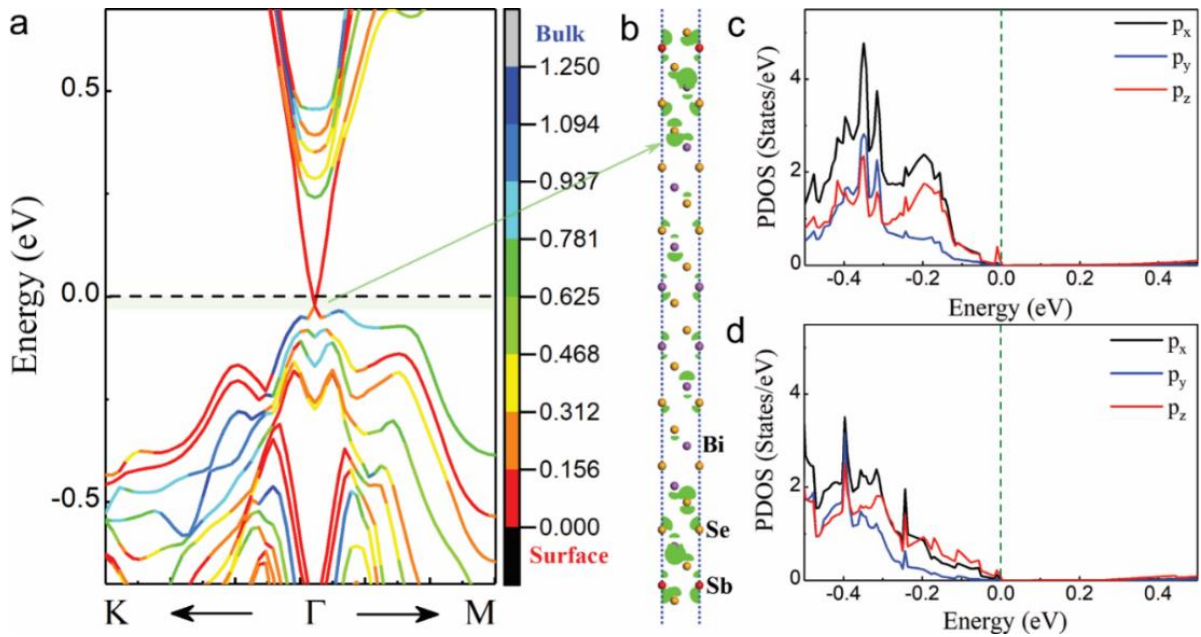

**Fig. 5.** Electronic structure of Sb doped $Bi_2Se_3$. (a) The projected band structure of the $Bi_{2-x}Sb_xSe_3$ surface slab, where the different colors denote the variation of band weight evolution between the surface band structure (red) and bulk band structure (blue). (b) The atomistic structure of the $Bi_{0.833}Sb_{0.167}Se_3$ surface slab superimposed with the partial charge density at the energy window as denoted in the green shadow in (a). The partial charge density is visualized with an iso-surface value of $5\times1.0^{-4}$ e/Å$^3$. The projected density of states on (c) the surface QLs and d) the central QLs of the $Bi_{2-x}Sb_xSe_3$ surface slab. The Fermi energy is shifted to 0 eV.

# Electronic correlation determining correlated plasmons in Sb-doped Bi$_2$Se$_3$

P. K. Das[1,*], T. J. Whitcher[1,2,†], M. Yang[4,§], X. Chi[1,3], Y. P. Feng[3], W. Lin[5], J. S. Chen[5], I. Vobornik[6], J. Fujii[6], K. A. Kokh[7,8], O.E. Tereshchenko[8,9], C. Z. Diao[1], Jisoo Moon[10], Seongshik Oh[10], A. H. Castro-Neto[2,3], M. B. H Breese[1,3], A. T. S. Wee[2,3], and A. Rusydi[1,2,3,11,12,‡]

[1] Singapore Synchrotron Light Source, National University of Singapore, 5 Research Link, 117603, Singapore

[2] Centre for Advanced 2D Materials, National University of Singapore, 6 Science Drive 2, 117546, Singapore

[3] Department of Physics, National University of Singapore, 2 Science Drive 3, 117576, Singapore

[4] Institute of Materials Research and Engineering, A*STAR (Agency for Science, Technology and Research), 2 Fusionopolis Way, 138634, Singapore

[5] Department of Materials Science and Engineering, National University of Singapore, 117575, Singapore

[6] Istituto Officina dei Materiali (IOM) - CNR, Laboratorio TASC, Area Science Park, S.S.14, Km 163.5, I-34149 Trieste, Italy

[7] V.S. Sobolev Institute of Geology and Mineralogy, Novosibirsk, 630090 Russia

[8] Novosibirsk State University, Novosibirsk, 630090 Russia

[9] A.V. Rzhanov Institute of Semiconductor Physics, Novosibirsk, 630090 Russia

[10] Department of Physics & Astronomy, Rutgers, The State University of New Jersey, 136 Frelinghuysen Road, Piscataway, New Jersey 08854, USA

[11] NUSSNI-NanoCore, National University of Singapore, 117576, Singapore

[12] NUS Graduate School for Integrative Sciences and Engineering, 117456, Singapore

Corresponding authors: [*]das@nus.edu.sg, [†]c2dwtj@nus.edu.sg, [‡]phyandri@nus.edu.sg

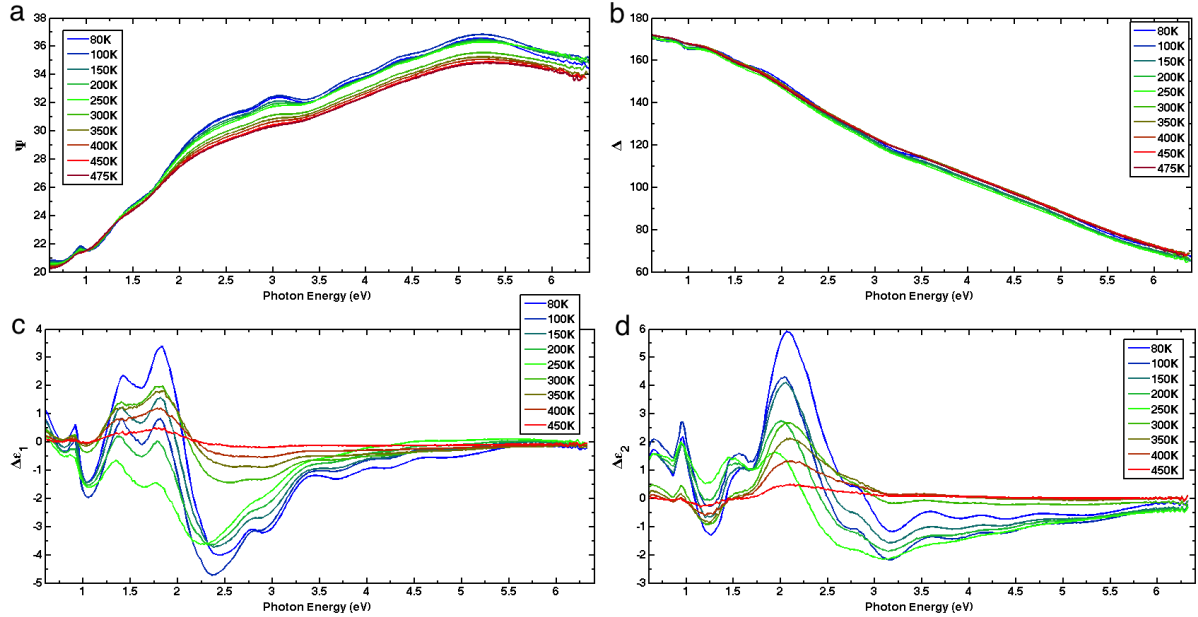

**Supplementary Fig. S1 | Spectroscopic Ellpsometry Data.** The output data of the spectroscopic ellipsometry measurements, (a) ψ and (b) Δ, the ratio of the amplitude and the difference in phase between the incident and reflected light, respectively. The difference in the (c) real and (d) imaginary parts of the dielectric function between 475 K and the colder temperature measurements.

Spectroscopic ellipsometry measurements were carried out using a variable-angle spectroscopic ellipsometer (V- VASE, J.A. Woollam Co.) with a rotating analyser and compensator at the Singapore Synchrotron Light Source (SSLS). The output data from the spectroscopic ellipsometry measurements are ψ and Δ, which are shown in Supplementary Figs. S1(a) and S1(b) respectively, as a function of temperature. The measurement ψ, is the ratio of the amplitudes of the polarised light incident on the target and the polarised light reflected from the target, whilst the measurement Δ, is the phase difference between the incident and reflected polarised light. Supplementary Fig. S1 also shows the difference in the real, Fig. S1(c), and imaginary, Fig. S1(d), parts of the complex dielectric function between the highest temperature of 475 K measured and the lower temperatures. There is clearly a very large change in the spectral weight as the sample is cooled with the biggest changes occurring around 2.0 eV, with other smaller changes at around 1.00 eV, 1.30 eV, 1.60 eV, and 3.20 eV.

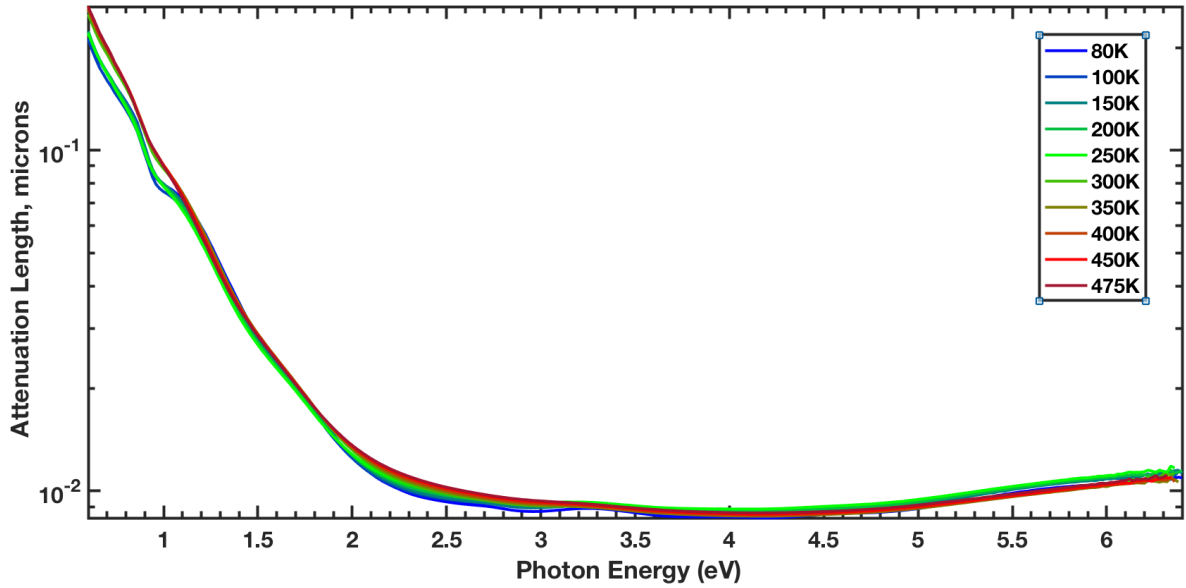

**Supplementary Fig. S2 | Attenuation Length of Sb-doped $Bi_2Se_3$.** The attenuation length of Sb-doped $Bi_2Se_3$ as a function of temperature from 0.6eV to 6.0eV.

The attenuation length of Sb-doped $Bi_2Se_3$ within the energy range of 0.6 – 6.0 eV at each temperature from 80 K to 475 K was calculated from the complex dielectric function shown in Fig. 2 of the main text. The attenuation length ranges from 250 nm at 0.6 eV down to around 10 nm above 6 eV.

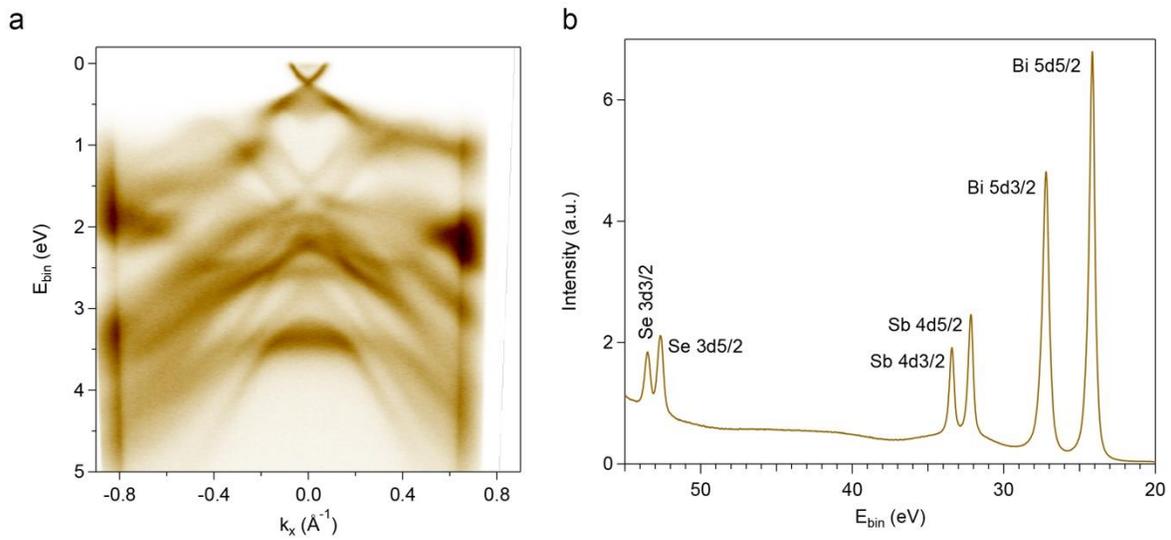

**Supplementary Fig. S3 | Deep valance bands and core levels.** (a) Deep valance band spectra presenting the details of band dispersion at deeper binding energies. A photon energy of $h\nu = 40$ eV was used to measure along the $\Gamma - K$ direction of Brillouin zone at 78 K. (b) Angular integrated shallow core levels of Bi, Sb, and Se are probed using 80 eV photon energy; confirming the Sb substitution in the system.

The deep valance bands down to a binding energy of 5 eV have also been measured. Supplementary Fig. S3(a) shows the band dispersion along the $\Gamma - K$ direction measured using a photon energy of 40 eV. The shallow core levels were also probed using a photon energy of 80 eV and are shown in Supplementary Fig. S3(b). We observe the core levels corresponding to Bi $5d_{5/2}, 5d_{3/2}$; Sb $4d_{5/2}, 4d_{3/2}$; and Se $3d_{5/2}, 3d_{3/2}$ transitions, respectively. This confirms the Sb substitution in this system.

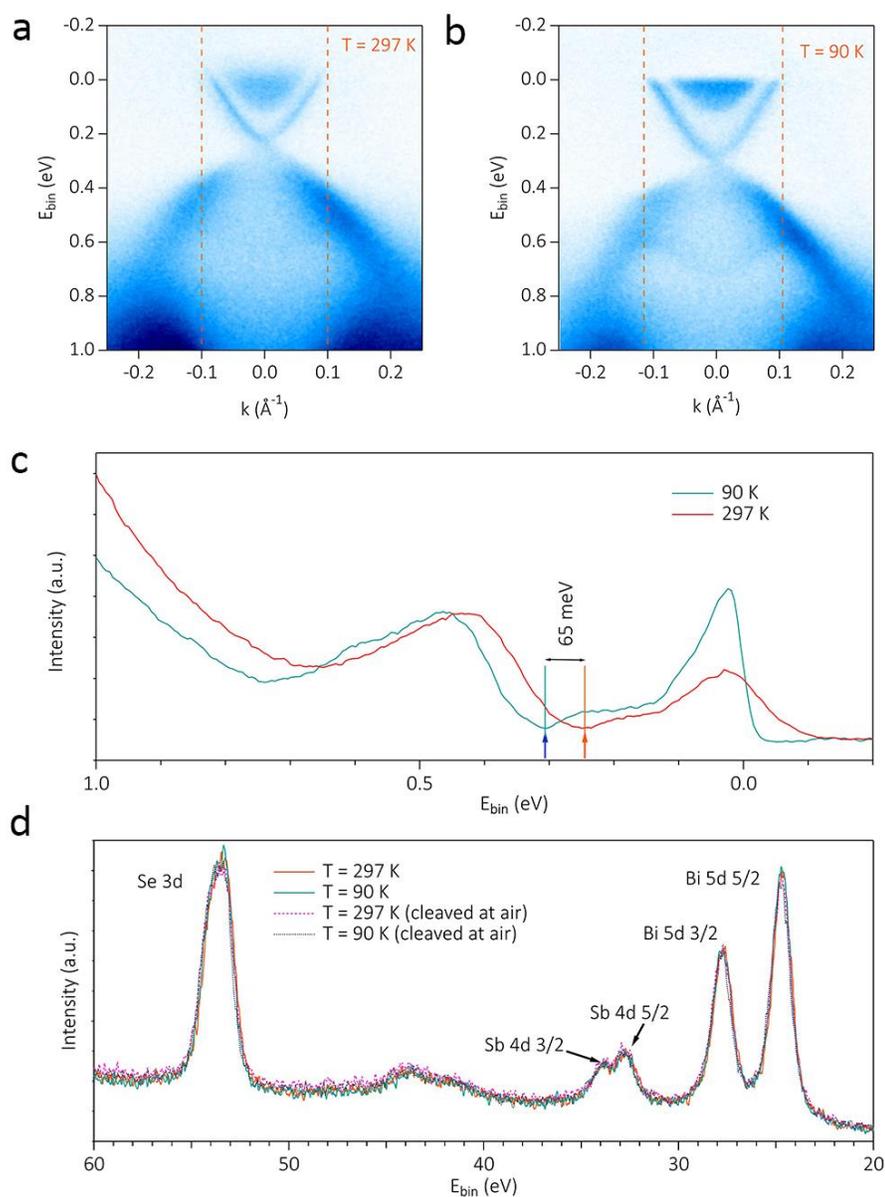

**Supplementary Fig. S4 | Temperature dependence ARPES and XPS study.** (a) Valence band spectra at room temperature and (b) at liquid nitrogen temperature (90 K). (c) Energy distribution curves (EDC) showing the shift in the Dirac point towards higher binding energy values at low temperature. The EDCs are extracted by integrating the intensity between the dotted line region as shown in panel a and b. (d) XPS spectra showing various Se, Sb, and Bi shallow core levels

measured at room temperature and low temperatures. The measurements were carried out on both in-situ and ex-situ (outside air) cleaved samples.

**Note on the effect of moisture and other environmental contamination:** In order to better understand the effect of temperature and moisture on the sample, we have recorded the ARPES spectra at both room temperature and liquid nitrogen temperature. We observed a shift of 65 meV in the Dirac point towards the higher binding energy at low temperature (See Supplementary Fig. S4 a-c). The shift in the valance band could be due to a combination of temperature, moisture [1] and other environmental effect, which introduces more n-type doping to the system.

In order to have a better ideal of the depth dependence, we have also performed XPS measurement at both high and low temperature using laboratory $Al\ K\alpha$ X-ray ($h\nu$ = 1486.6 eV), which is relatively more bulk sensitive than the UV light ($h\nu$ = 21.21 eV) used for ARPES experiment. We have measured the XPS spectra for both in-situ cleaved and outside air cleaved samples. We do not see any significant changes in the peak height ratio of Bismuth, Antimony and Selenium. The results are presented in Supplementary Fig. S4 d. We only observed the characteristic Bi $5d_{5/2}$ and Bi $5d_{3/2}$ doublet peaks corresponding to Bi-Se bonding. We do not see any signature of Bi bilayer peak appearing in our samples as observed by M. T. Edmonds et al. [2] in pure $Bi_2Se_3$ system after atmospheric exposure. The authors also reported large reduction in the Se $3d$ core level peak due to Bismuth bilayer formation, which we do not observed in case of our Sb-doped $Bi_2Se_3$ samples in our experimental conditions. Furthermore, the authors reported several hundred meV shift in the peak position, which was not seen in our case. We have not observed any signature of the sample surface getting oxidized in our experimental conditions [3]. We believe that the changes in the valence band (ARPES) spectra occurs due the changes in the immediate surface vicinity, while the bulk of the sample remains unaffected by the moisture or another environmental contamination. As UV ARPES is extremely surface sensitive, the surface contamination effects are expected to show up only in the ARPES data.

On the other hand, spectroscopis ellipsometry technique is not as surface sensitive as photoelectron spectroscopy techniques, where the light penetration depth is tens or even hundreds of nm (c.f. **Supplementary Fig. S2**). It is noted that the effect of moisture in shift the ARPES spectra is a function of time, longer time in the cold manipulator introduces gradual shift in the spectra which eventually stabilizes. The ellipsometry spectra are not function of time and reproducible over different heating/cooling cycles. We believe that the changes in the optical spectra are solely due to the temperature, and not due to the minor changes in the immediate vicinity of the surface caused by the moisture.

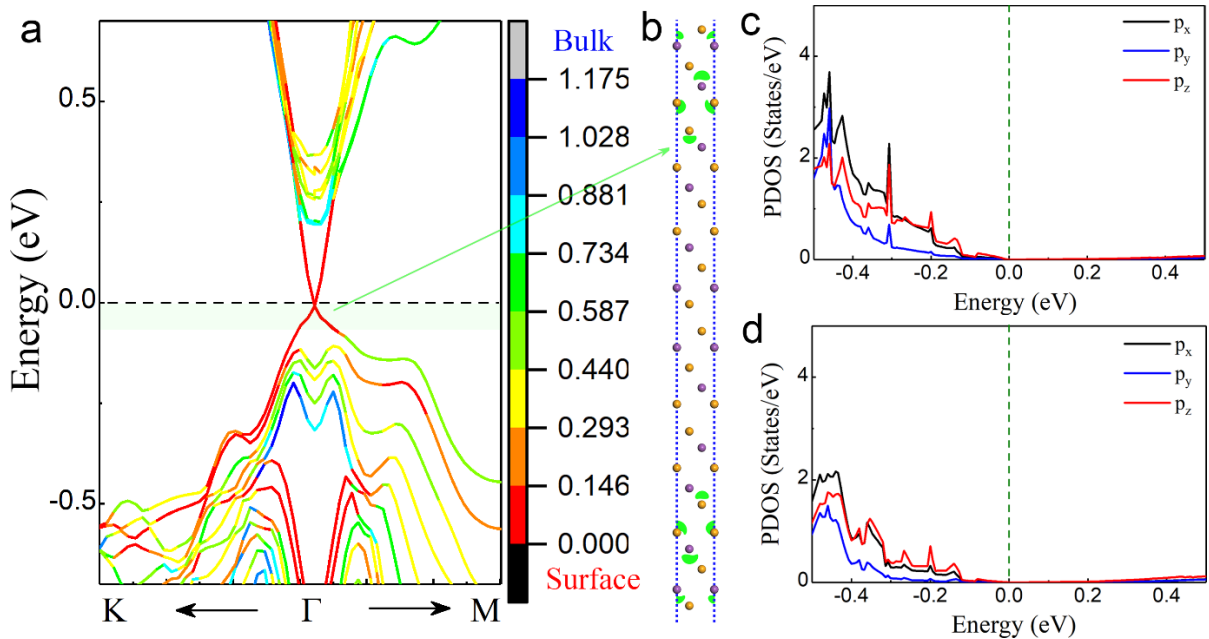

**Supplementary Fig. S5 | Electronic structure of $Bi_2Se_3$.** (a) The projected band structure of the $Bi_2Se_3$ surface slab with 6 QL layers, where the different colour denotes the variation of band

weight evolution between surface band structure (red colour) and bulk band structure (blue colour). (b) The atomistic structure of the $Bi_2Se_3$ surface slab superimposed with the partial charge density at the energy window as denoted in the green shadow in (a). The partial charge density is visualized with an iso-surface value of $5\times1.0^{-4}$ e/bohr$^3$. The projected density of states on (c) the surface QLs and (d) the central QLs of $Bi_2Se_3$ surface slab. The Fermi energy is shifted to 0 eV.

First principle calculations have been carried out in order to simulate the complex band structure of the $Bi_{2-x}Sb_xSe_3$ and $Bi_2Se_3$ topological insulators. These calculations can be used to identify changes in the band structure from Sb doping and by comparing with ARPES results. Details of the calculations are given in the main text. Supplementary Fig. S5(a) shows the calculated electronic structure bands of $Bi_2Se_3$ with the colour of the band representing the contribution from the QL layers as a function of distance from the surface. For example, bands that are red come from QL layers that are close to the surface whilst bands that are blue come from QL layers located within the bulk. It is clear that the contribution to most of the bands surrounding the Fermi level and Dirac point come from the surface layers, which is in contrast to the $Bi_{2-x}Sb_xSe_3$ electronic structure seen in Fig. 5 of the main text where the bulk layers have a much more significant contribution to the lower Dirac cone.

Supplementary Fig. S5(b) shows the atomistic structure of the 6 QL layer $Bi_2Se_3$ surface slab with the green areas highlighting the origin of the partial charge density that contributes to the states closest to the Fermi level [shown as the green shadow in Supplementary Fig. S5(a)]. All of the contributions to the surface states come from the surface layers whereas in $Bi_{2-x}Sb_xSe_3$ there are more contributions from the bulk QL layers as seen in Fig. 5(b). Supplementary Fig. S5(c) and S5(d), respectively. The DOSs closest to the Fermi level are dominated by the $p_z$ orbitals of the Bi atoms from the surface QLs but again there is very little contribution from any of the orbitals of the bulk Bi atoms.

**References.**